\documentclass[aps,prd, floatfix,twocolumn,preprintnumbers,superscriptaddress,showpacs]{revtex4-1}
\usepackage{graphicx}
\usepackage{slashed}
\usepackage{amsmath,amsfonts,bm,bbold}
\usepackage{amssymb}
\begin{document}


\author{A.~V.~Radyushkin}
\affiliation{Physics Department, Old Dominion University, Norfolk,
             VA 23529, USA}
\affiliation{Thomas Jefferson National Accelerator Facility,
              Newport News, VA 23606, USA
}
\affiliation{Bogoliubov Laboratory of Theoretical Physics, JINR, Dubna, Russian
             Federation}

\title{Modeling Nucleon Generalized Parton Distributions  }

\begin{abstract}

We discuss  
 building models for  nucleon generalized parton distributions 
 (GPDs)  $H$ and $E$ that are based on the formalism   of double distributions (DDs).  We found that the usual 
 ``DD+D-term'' construction should be amended by an extra term, 
 $\xi E_+^1 (x,\xi)$  built from the $\alpha/\beta$ moment of the
 DD $e(\beta,\alpha)$  that generates    GPD $E(x,\xi)$.
Unlike  the $D$-term, this function has support in the whole
$-1 \leq x \leq 1$ region, and in general does not vanish at the border points
$|x|=\xi$.


\end{abstract}

\pacs{11.10.-z,12.38.-t,13.60.Fz}
\maketitle

   
\section{Introduction}

The   studies of Generalized Parton Distributions
(GPDs)  \cite{Mueller:1998fv,Ji:1996ek,Radyushkin:1996nd,Collins:1996fb}
require building theoretical   models for  GPDs  which  satisfy several nontrivial requirements  
such as polynomiality \cite{Ji:1998pc},  positivity \cite{Martin:1997wy,Pire:1998nw,Radyushkin:1998es},
  hermiticity  \cite{Mueller:1998fv}, time reversal invariance \cite{Ji:1998pc}, etc.
 The constraints  follow from the most  general principles  of  quantum  field  theory. 
Polynomiality (that may be traced back to Lorentz  invariance)
  imposes   the restriction  that 
$x^n$ moment of a GPD $H(x,\xi;t)$  must be  a polynomial in $\xi$ of the
order not  higher than $n+1$. 
This  property is  automatically obeyed by 
GPDs  constructed from Double Distributions (DDs) 
 \cite{Mueller:1998fv,Radyushkin:1996nd,Radyushkin:1996ru,Radyushkin:1998es}. 
(Another way to impose the 
 polynomiality   condition  onto model GPDs is ``dual  parameterization''
 \cite{Polyakov:2002wz,Polyakov:2007rw,Polyakov:2007rv,SemenovTianShansky:2008mp,Polyakov:2008aa}).  
Thus, within the DD approach,  the problem of  constructing a  model  for a  GPD 
 converts into a  problem of  building a  model
 for the relevant DD.
 
Double distributions 
 $F(\beta, \alpha;t)$   behave  like 
 usual parton distribution functions (PDFs)  with respect to 
 its variable $\beta$,  as a meson 
 distribution amplitude  (DA) with respect to $\alpha$, and as  a form  factor 
 with respect to the invariant  momentum transfer $t$.
 The {\it factorized DD ansatz} (FDDA) 
\cite{Radyushkin:1998es,Radyushkin:1998bz} proposes 
to  build a  model DD $F(\beta, \alpha)$  
(in the simplified formal $t=0$  limit) as  a product of the usual 
parton density  $f(\beta)$
and a  profile function $h(\beta, \alpha)$  that  has an $\alpha$-shape of a 
meson DA. However,  it  was noticed  \cite{Polyakov:1999gs}
that in  the case of isosinglet pion GPDs, FDDA 
does not produce the highest, $(n+1)^{\rm st}$  
power of $\xi$ in the $x^n$ moment of $H(x,\xi)$.
To cure this problem, a  ``two-DD''  parameterization for pion GPDs was proposed
\cite{Polyakov:1999gs}, with the second DD $G(\beta, \alpha)$
capable of generating,  among others,  the required $\xi^{n+1}$ power. 
It was also proposed \cite{Polyakov:1999gs} to use  a  ``DD plus D'' parameterization
in which the second DD $G(\beta, \alpha)$  is reduced to a function $D(\alpha)$
of one variable,  the $D$-term , that is   solely  responsible for the $\xi^{n+1}$ contribution.
As  emphasized in Ref. \cite{Polyakov:1999gs}, one should also
add $D$-term in  case of nucleon distributions.
The importance of the $D$-term and  its  physical  interpretation
were  studied in further works (see Ref.~\cite{Goeke:2001tz}  and references therein).

 In the pion  case, it was shown \cite{Teryaev:2001qm}  
 that one can reshuffle 
 terms between $F$ and $G$ functions  of 
 the $F+G$ decomposition without changing the sum
 (``gauge invariance''). 
 Furthermore, 
 it was  found in Ref. \cite{Belitsky:2001ns},
 that  one can  write 
a  parameterization 
that   incorporates just one function $f(\beta,\alpha)$,
 but still produces 
all the required powers up to $\xi^{n+1}$.
A model for the pion GPD based on this representation
was built in our paper  \cite{Radyushkin:2011dh}. 
An important ingredient of our  construction 
was  separation of DD $f(\beta,\alpha)$ in its 
``plus'' part $[f(\beta,\alpha)]_+$
that gives zero after integration over $\beta$,
and $D$-term part $\delta (\beta) D(\alpha)/\alpha$. 
For DDs  singular in small-$\beta$ region, such a separation
serves also as a renormalization prescription
substituting a formally divergent integral over $\beta$
by ``observable'' $D$-term.

In the present paper, we apply the technique of Ref.
\cite{Radyushkin:2011dh} 
(see also \cite{Radyushkin:2012gba})
for building models of nucleon GPDs
$H(x,\xi)$ and $E(x,\xi)$. 
The paper is organized as follows.
To make it self-contained, we start, in Sect. II,  with 
a short review of the basic facts about DDs, GPDs and $D$-term,
using a toy model with scalar quarks,
that allows to  illustrate  essential features of GPD theory  
avoiding  complications related to spin. 
In Sect. III, we describe  the theory of  pion GPD $H(x,\xi)$,
presenting the results of 
 Ref. \cite{Radyushkin:2011dh} in a form suitable 
 for generalization onto the nucleon case. 
In Sect. IV, we recall the basic ideas of the
factorized DD Ansatz  of Refs. \cite{Radyushkin:1998es,Radyushkin:1998bz}.
In Sect. V, we use the formalism described in previous sections
for building DD models for nucleon GPDs $H(x,\xi)$
and $E(x,\xi)$. 

An essential   point is that two 
functions $A$ and $B$ associated with two basic 
Dirac structures present in the twist decomposition
of the nucleon matrix element 
do not coincide with $H$ and $E$.
In fact, $A=H+E$ and $B=-E$.
What is most important, $A$ and $B$ have different
types of DD representation:
$A$ is given by the simplest (scalar-type) DD representation,
while $B$ is given by a more complicated 
representation coinciding with the one-DD
parametrization of the pion case. 
Thus, building a model for $H$ one should deal with  a sum 
$A+B$,  the terms of which have 
different-type DD representations.
The result of this mismatch is a term,
which we call $\xi E_+^1 (x,\xi)$ that 
is given by the ``plus'' part of the $\alpha/\beta$ 
moment of DD $e(\beta, \alpha)$ 
used in parametrization for $E(x,\xi)$ GPD. 
The term  $\xi E_+^1 (x,\xi) $ should be included  in the model 
for GPD $H(x,\xi)$. 
However, unlike the $D$-term contribution,
the function  $\xi E_+^1 (x,\xi) $ in general does not vanish 
both at the border points $|x|=\xi$ and also 
outside the central region $|x|\leq \xi$.

In final  section, we summarize the results of the paper.

\section{Basics of theory for DDs and   GPDs}

\subsection{Matrix elements and DDs}

Parton distributions provide  a convenient way to parametrize 
 matrix elements of local operators  that accumulate information about 
 hadronic structure. 
Various  types of distributions differ by the nature of the matrix elements involved.
 In particular, to define GPDs, one starts with non-forward matrix elements
\mbox{$\langle P+r/2 |  \ldots |P-r/2 \rangle $,} with $P$ being the 
average  of the  initial and final hadron momenta,
and $r$ being their  difference.   
In scalar case (which illustrates many essential features 
without irrelevant complications)  we have 
\begin{align}
&  \langle P+r/2 |   \psi(0) \{\stackrel{\leftrightarrow}{\partial}_{\mu_1} 
\ldots  \stackrel{\leftrightarrow}{\partial}_{\mu_n}\} \psi (0)|P-r/2 \rangle  \nonumber \\  &=
\sum_{l=0}^{n-1}A_{nl}\{P_{\mu_1}\ldots  P_{\mu_{n-l}}
r_{\mu_{n-l+1}}  \ldots   r_{\mu_n}\}\nonumber \\  &+  A_{nn}  \{r_{\mu_{1}}  \ldots   r_{\mu_n}\}   \  .
\label{scalarOn}
\end{align}
The notation $\{ \ldots \}$ indicates the symmetric-traceless part
of the enclosed tensor.  Since two vectors are involved,
we have $n+1$ distinct tensor structures differing in the number  $l$
of  $r$  factors involved. In the forward $r=0$ limit, only the 
$A_{n_0}$ coefficients are visible. Another extreme case is 
$l=n$,   corresponding to the tensor $\{r_{\mu_{1}}  \ldots   r_{\mu_n}\}$
built solely from the $r$ momentum. 

The forward $r=0$ limit corresponds to matrix elements 
defining usual parton distributions $f(x)$ as a function whose 
moments produce $A_{n0}$:
\begin{align}
\int_{-1}^1 f(x) x^n \, dx = A_{n_0} \ .
\end{align}
The  parton interpretation of $f(x)$ is that 
it describes a parton with momentum $xP$. 
This definition of $f(x)$ may be rewritten in terms of 
matrix elements of operators on the light cone:
\begin{align}
& \langle P |   \psi(-z/2) \psi (z/2)|P \rangle  \nonumber \\  &
=  
\int_{-1}^1 f(x)  \,  e^{-i x(Pz) } \, dx \,
 +  {\cal O} (z^2) \   . 
\label{PDF}
\end{align}

In a general non-forward case, the parton carries
the fractions of both $P$ and $r$ momenta.  
Note, that in  the momentum  representation,   the  derivative $ \stackrel{\leftrightarrow}{\partial}_{\mu}\ $
converts  into  the   average $\bar k _\mu =(k_\mu+k'_\mu)/2$ of  the   initial  $k$  and final $k'$ quark  
momenta. After integration over $k$, $(\bar k)^n$ should produce the $P$  and $r$  factors
in the r.h.s.  of \mbox{ Eq. (\ref{scalarOn}).}  In this   sense, one may treat $(\bar k)^n$
as $(\beta P + \alpha r/2)^n$  and   define the {\it double distribution}  (DD)
 \cite{Mueller:1998fv,Radyushkin:1996nd,Radyushkin:1996ru,Radyushkin:1998es} 
\begin{align}
 \frac{n!}{(n-l)! \,  l! \, 2^{l}} \int_{\Omega}  F(\beta, \alpha) \beta^{n-l} \alpha^l \, d\beta \, d\alpha=
A_{nl}
\label{Anl}
\end{align}
as a function whose $\beta^{n-l} \alpha^l$   moments are  proportional to the coefficients
$A_{nl}$.   It   can  be  shown  \cite{Mueller:1998fv,Radyushkin:1996nd,Radyushkin:1998bz} that the support region
$\Omega$  is given by the rhombus $|\alpha|+|\beta| \leq 1$.
These   definitions result in the \mbox{``DD   parameterization''} 
\begin{align}
& \langle P-r/2 |   \psi(-z/2) \psi (z/2)|P+r/2 \rangle  \nonumber \\  &
=  
\int_{\Omega}  F(\beta, \alpha) \,  e^{-i \beta (Pz) -i\alpha (rz)/2} \, d\beta \, d\alpha
 +  {\cal O} (z^2) \   . 
\label{DDF}
\end{align}
 of the
matrix element. 

\subsection{Introducing GPDs  and $D$-term}

Another parametrization of the non-forward matrix element
is in terms of {\it generalized parton distributions}. 
In scalar case
GPDs are  defined by   
\begin{align}
&  \langle P-r/2 |   \psi(-z/2) \psi (z/2)|P+r/2 \rangle \nonumber \\  & =  
\int_{-1}^1 \,  e^{-ix(Pz) } H(x,\xi) \, dx +  {\cal O} (z^2) \   , 
\label{GPDdef}
\end{align}
and relation between GPD and DD functions is given by 
\begin{align}
& H(x,\xi) =  
\int_{\Omega}  F(\beta, \alpha) \,  \delta (x - \beta -\xi \alpha)  \, d\beta \, d\alpha  \  .
\label{GPD}
\end{align}
The skewness parameter $\xi$  in this definition corresponds to the ratio
$(rz)/2(Pz)$.

In the forward limit  $\xi=0$, GPD  $H(x,\xi)$ converts into the usual 
parton distribution $f(x)$. Using DDs, we may write
\begin{align}
f(x) &=
     \int_{-1+|x|} ^{1  -|x|}  F(x,\alpha) \,   d\alpha   \   .
\label{GPDtofx}
\end{align}
Thus,  the forward  distributions $f(x)$  are  obtained by integrating DDs
over vertical lines $\beta =x$  in the $(\beta,\alpha)$ plane.
As discussed above,  $f(x)$ is defined through the coefficients
$A_{n0}$ corresponding to tensors without $r$ factors.
Similarly, one can treat the $A_{nn}$ coefficients,
corresponding to tensors without $P$ factors,  as the moments
of another  function $D(\alpha)$
\begin{align}
\int_{-1}^1  D(\alpha)  \,  (\alpha/2)^n \, d\alpha= A_{nn}  \  ,
\end{align}
the {\it $D$-term}  \cite{Polyakov:1999gs}.  
From the  definition of DD (\ref{Anl}), it follows that
\begin{align}
D(\alpha) =   \int_{-1+|\alpha|}^{1-|\alpha|}  F(\beta,\alpha)\,  d \beta  \  ,
\label{D}
\end{align} 
i.e., $D$-term $D(\alpha)$  is obtained from DD  $F(\beta,\alpha)$ by integration over 
horizontal lines in the $\{\beta,\alpha\}$ plane. 
In this sense, one can think of ``vertical'' projection of DD that
produces the forward distribution $f(\beta)$, and ``horizontal''
projection that produces $D$-term $D(\alpha)$. 

Taking the $x^n$  moment of GPD $H(x,\xi)$
\begin{align}
\int_{-1}^1 \, H(x,\xi) \, x^n \, dx = \sum_{l=0}^{n}  A_{nl} (2 \xi)^l \ , 
\label{GPDdef}
\end{align}
we see that the coefficients $A_{nn}$ are responsible for the highest 
power of skewness $\xi$ in this expansion.

\subsection{DD plus D parametrization}

Parameterizing the matrix element (\ref{scalarOn}), one may wish
to separate the $A_{nn} $  terms 
that are accompanied by tensors built 
from the momentum transfer vector  $r$  only, 
and,  thus,  are  invisible in the forward $r=0$ limit,
i.e., to separate the $D$-term contribution. 
This  
can be made   by simply using 
\begin{align}
e^{-i\beta(Pz)}= [e^{-i\beta (Pz)}-1] +1
\end{align}
which converts  the DD-parameterization 
into a \mbox{``DD$_+$ plus D''}  parameterization
\begin{align}
&  \langle P-r/2 |   \psi(-z/2) \psi (z/2)|P+r/2 \rangle  \nonumber \\  &
=  
\int_{\Omega}  [F(\beta, \alpha)]_+ \,  e^{-i \beta (Pz) -i\alpha (rz)/2} \, d\beta \, d\alpha
 \nonumber \\  &
+  \int_{-1}^1  D(\alpha)  \,  e^{-i\alpha (rz) /2} \, d\alpha  +  {\cal O} (z^2) \   ,
\label{DDplusD}
\end{align}
where
\begin{align}
[F(\beta, \alpha)]_+  = 
 F(\beta, \alpha) -\delta (\beta)
   \int_{-1+|\alpha|}^{1-|\alpha|}
F(\gamma,\alpha)\,  d \gamma 
\end{align}
is the DD with subtracted $D$-term
 given by  Eq.(\ref{D}). 
Then
\begin{align}
 F(\beta, \alpha)  = [F(\beta, \alpha)]_+  + 
\delta (\beta)
D (\alpha)  
\end{align}
and 
\begin{align}
H(x, \xi) =  H_+ (x, \xi) + \frac{D(x/\xi)}{|\xi|} \  ,
\end{align}
where 
\begin{align}
H_+(x,\xi) = & 
\int_{\Omega}  [F(\beta, \alpha) ]_+ \,  \delta (x - \beta -\xi \alpha)  \, d\beta \, d\alpha  \nonumber \\ & = 
\int_{\Omega}  F(\beta, \alpha) \, \Big [  \delta (x - \beta -\xi \alpha)  \nonumber \\ & \hspace{2cm}  
-  \delta (x -\xi \alpha) \Big ]  \, d\beta \, d\alpha  
\label{GPD}
\end{align}
is the ``plus'' part of GPD $H(x,\xi) $.

A straightforward   observation is   that  the 
 $x^n$ moment of $H_+ (x,\xi)$  does not contain 
  the highest, namely the $n^{\rm th}$ power of $\xi$, 
  since the relevant   integral
\begin{align}
  \int_{\Omega}  
                          \alpha^n \,  \left [  F (\beta,\alpha) \right ]_+ \, 
                                    \, d\beta \, d\alpha
 \end{align}
 vanishes because the integrand is a ``plus'' distribution 
 with respect to $\beta$. 

For  $n=0$,   the highest  power is $\xi^0$, and since  the \mbox{$n=0$}  moment of $H_+ (x,\xi)$ should   not contain this highest  power,  
it contains no powers of $\xi$  at all, i.e. it vanishes:
\begin{align}
\int_{-1}^1 \, H_+(x,\xi) \, dx = \int_{\Omega}  
                           \left [  F (\beta,\alpha) \right ]_+ \, 
                                    \, d\beta \, d\alpha = 0 \ 
 . 
\label{H+zero}
\end{align}
Thus, $ H_+(x,\xi)$ has the same property with respect to integration
over $x$ as a ``plus'' distribution
\begin{align}
[h(x)]_+ = h(x) - \delta (x) \int_{-1}^1  h(z) dz \  .
\end{align}
However,   $ H_+(x,\xi)$  may be  a pretty smooth function,
without any $\delta (x)$ terms. It should just possess 
regions of positive and negative values of $ H_+(x,\xi)$  
averaging to zero after $x$-integration. 

\subsection{$D$-term as a separate entity}

 In the simple model with scalar quarks, 
  one  may  just  use  the original DD  $F(\beta,\alpha)$
  without splitting it into the ``plus'' part and the $D$-term.  
  One may imagine that the DD $F (\beta,\alpha) $ 
  is some smooth function on the rhombus,
  with nothing spectacular happening on the $\beta =0$
  line. In such a case, one may, of course,  write 
  $ F(\beta, \alpha)  = [F(\beta, \alpha)]_+  + 
  \delta (\beta) D(\alpha)$, with the $D$-term accompanied 
  by the $\delta (\beta)$ function, but this term is precisely canceled 
  by the $\sim \delta (\beta)$ term contained in 
  $ [F(\beta, \alpha)]_+ $.  
  
  However, if the theory 
  allows purely $t$-channel exchanges, 
  then the relevant diagrams  generate 
  $\sim \delta (\beta)$ terms not necessarily  connected 
  to  other contributions. E.g.,  our scalar quarks  may 
  have a quartic interaction, and  the $t$-channel 
  loop would generate a $ \delta (\beta) \varphi (\alpha)$
  type 
  contribution into $F(\beta,\alpha)$.

 Furthermore, $D$ term is  formally given by  
 the  integral of $F(\beta,\alpha)$. An implicit assumption
 is that this integral converges, which is the case if  
 $F(\beta, \alpha)$ is not too singular.  
 Note, however, that the integral of $F(\beta, \alpha)$ 
 over $\alpha$ gives $f(\beta)$,
 a usual parton distribution  which 
 are known to have a  singular  $\sim \beta^{-a}$ 
 behavior  for small $\beta$.
This means that the $\beta$-profile of DD $F(\beta, \alpha)$ 
should  be similar to that of $f(\beta)$, and also be  singular 
 in the  $\beta \to 0 $ region,    $F(\beta, \alpha)\sim \beta^{-a}$. 
 The integral over $\beta$ converges if $a<1$.
 However, as we will see in Sec. III B, one may need  the  
 integrals involving $F(\beta,\alpha)/\beta$ which diverge 
 for any positive $a$.  
 The integral for $[F(\beta,\alpha)]_+$ 
 still converges for  $a<1$, and 
  the role of the $D$ term in this  case 
   is to substitute   the divergent integral
   \begin{align}
    \int_{-1+|\alpha|}^{1-|\alpha|}  F(\beta,\alpha)\,  d \beta  
   \end{align} 
   by a finite function $D(\alpha)$
   whose $\alpha^n$ moments then give 
   finite coefficients $A_{nn}$.
   In this case, the ``DD$_+$  plus D'' separation 
   serves as a renormalization prescription 
   defining the moments of DD. 
 
 An attempt to
 consistently  ``implant''  the Regge 
 behavior into a quantum field theory construction
 was made in Ref. \cite{Szczepaniak:2007af},
 where  a dispersion relation   was used  for 
 an amplitude that has  $s^a$  Regge behavior at  large
 energies. For any positive $a$, such a relation
 requires a subtraction, which (as shown in Refs.\cite{Szczepaniak:2007af,Radyushkin:2011dh})
  results in a 
 $\delta (\beta)\varphi (\alpha)$ term contributing to   $D(\alpha)$.

\section{Pion DDs and GPDs}

\subsection{Two-DD representation}

  In fact, $D$-term was introduced first 
  \cite{Polyakov:1999gs} in the context of pion 
  GPDs, with pion made  of spinor quarks.
  In that case,  
 it is  more difficult to avoid 
an  explicit introduction  of the   $D$-term as an extra function. 
The basic  reason    is  that the matrix element   of the bilocal  operator 
in pion   case has two parts
 \begin{align}
&  \langle P-r/2 |   \bar \psi(-z/2) \gamma_\mu \psi (z/2)|P+r/2 \rangle  |_{\rm twist-2}
 \nonumber \\  &
=  2P_\mu f \bigl ((Pz),(rz),z^2 \bigr )  + r_\mu g\bigl ((Pz),(rz),z^2\bigr )   \  .
\label{pmurmu}
\end{align}
This  suggests  a   parametrization  with  two DDs
corresponding to $f$   and $g$ functions  \cite{Polyakov:1999gs}.  
For the matrix element   (\ref{pmurmu})
multiplied by $z^\mu$  
(the object  one obtains doing  the  leading-twist 
factorization  for the Compton amplitude \cite{Balitsky:1987bk} )   this gives 
 \begin{align}
& z^\mu  \langle P-r/2 |   \bar \psi(-z/2) \gamma_\mu \psi (z/2)|P+r/2 \rangle 
\nonumber 
\\  &
=  
\int_{\Omega}   e^{-i \beta (Pz) -i\alpha (rz)/2} \,\biggl [ 2(Pz)  F(\beta, \alpha) 
\nonumber 
\\  & + (rz)   G(\beta, \alpha) \biggr ]   \, d\beta \, d\alpha \ 
+{\cal O} (z^2)  .  
\label{twoDD}
\end{align}
Then GPDs are  given by a DD representation 
\begin{align}
& H(x,\xi) =  
\int_{\Omega}  \bigl [F(\beta, \alpha) +\xi G(\beta,\alpha) \bigr ] \,  \delta (x - \beta -\xi \alpha)  \, d\beta \, d\alpha  \  ,
\label{GPDFG}
\end{align}
that involves two DDs: $F(\beta, \alpha) $ and $ G(\beta,\alpha) $.
The highest power  $\xi^{n+1}$ for the $x^n$ moment of $H(x,\xi)$
is given now by the $G$ term, which one can 
separate  
\begin{align}
 G(\beta, \alpha)  = [G(\beta, \alpha)]_+  + 
\delta (\beta)
D (\alpha)  
\end{align}
into a ``plus'' part  and $D$-term 
\begin{align}
D(\alpha) =   \int_{-1+|\alpha|}^{1-|\alpha|}  G(\beta,\alpha)\,  d \beta  \  .
\end{align} 
As a result, 
\begin{align}
H(x, \xi) =  F (x, \xi)  + \xi G_+ (x, \xi) + {\rm sgn} (\xi) {D(x/\xi)} \  ,
\end{align}
where
\begin{align}
& F(x,\xi) =  
\int_{\Omega}  F(\beta, \alpha) \,  \delta (x - \beta -\xi \alpha)  \, d\beta \, d\alpha  
\label{GPD}
\end{align}
and
\begin{align}
G_+(x,\xi) =  
\int_{\Omega}  G(\beta, \alpha) \, \Big [ & \delta (x - \beta -\xi \alpha)  \nonumber \\ &
-  \delta (x -\xi \alpha) \Big ]  \, d\beta \, d\alpha   \   .
\label{GPD}
\end{align}
The forward distribution $f(x)$ in two-DD formulation is obtained from the DD $F$  only:
\begin{align}
f(\beta) &=
     \int_{-1+|\beta|} ^{1  -|\beta|}  F(\beta,\alpha) \,   d\alpha   \   .
\label{Ftofx}
\end{align}
Thus, $D$-term and $f(x)$ are obtained from different
functions, so the $D$-term is indeed  looking 
  like an independent
entity.

\subsection{One-DD representation}

Note that  the Dirac  index $\mu$  is symmetrized in the  local twist-two operators
\mbox{$\bar \psi \{\gamma_\mu  \stackrel{\leftrightarrow}{\partial}_{\mu_1} \ldots  \stackrel{\leftrightarrow}{\partial}_{\mu_n}\} \psi$}
with the $\mu_{i}$ indices related to the derivatives. Thus, 
one may expect that it also produces the factor  $\beta P_{\mu}+ \alpha r_{\mu}/2$.
In  Ref.~\cite{Belitsky:2000vk}, it was shown   that this is really the case.
In other words, not only the exponential produces the $z$-dependence  in the
combination $\beta (Pz) +\alpha (rz)/2$,
but also the pre-exponential terms  come  in the 
\mbox{$\beta (Pz) +\alpha (rz)/2$}  combination. The  result is 
a representation in which 
\begin{align}
 &2(Pz)  F(\beta, \alpha)  + (rz)   G(\beta, \alpha) \nonumber \\  & = [ 2\beta (Pz) +\alpha (rz)]
 f(\beta, \alpha)  \   , 
\end{align}
that  corresponds to 
$$F(\beta,\alpha) =\beta f(\beta,\alpha)$$
 and   $$G(\beta,\alpha) = \alpha f(\beta,\alpha)\  .$$
Thus,  one deals formally with just one DD  $f(\beta,\alpha)$. 
 The two-DD representation for GPDs 
 (\ref{GPDFG}) 
converts into 
\begin{align}
H(x,\xi) &=  
\int_{\Omega}  (\beta  +\xi  \alpha)  f(\beta,\alpha) \,  \delta (x - \beta -\xi \alpha)  \, d\beta \, d\alpha
\nonumber \\
&= x \int_{\Omega}    f(\beta,\alpha) \,  \delta (x - \beta -\xi \alpha)  \, d\beta \, d\alpha  
\label{GPDf}
\end{align}
in the ``one-DD'' formulation.

The  $D$-term  in the one-DD case   is given by
\begin{align}
D(\alpha) = \alpha \int_{-1+|\alpha|} ^{1  -|\alpha|} f(\beta,\alpha) \, d\beta  \   ,
\label{DTsingle}
\end{align}
and one may  write $f(\beta,\alpha)$ as a sum 
\begin{align}
  f(\beta,\alpha)= [f(\beta,\alpha)]_+  +\delta (\beta) \frac{D(\alpha)}{\alpha}  
\end{align}
of its ``plus'' part
\begin{align}
 [f(\beta,\alpha)]_+=   f(\beta,\alpha) - \delta (\beta)  
  \int_{-1+|\alpha|} ^{1  -|\alpha|} f(\gamma,\alpha) \, d\gamma
  \label{Plussingle}
\end{align}
and $D$-term part $ \delta (\beta) {D(\alpha)}/{\alpha}$.

For the GPD $H(x,\xi)$, the ``DD$_+$+ D'' separation corresponds 
to the representation 
\begin{align}
H(x,\xi)\equiv H_+(x,\xi)+{\rm sgn}(\xi) D(x/\xi) \ , 
\label{HplusD}
\end{align}
where 
\begin{align}
 \frac{H_+(x,\xi) }{x}  \equiv 
 \int_{\Omega} f(\beta,\alpha)   \,\Big [ &  \delta (x - \beta -\xi \alpha) 
 \nonumber \\ &
 - \delta (x - \xi \alpha )  \Big ]
  \, d\beta \, d\alpha   \   .
\label{GPDfplus}
\end{align}
Using 
$f(\beta,\alpha) = F(\beta,\alpha)/\beta$
we may rewrite 
\begin{align}
& H(x,\xi) =  
\int_{\Omega}  (\beta  +\xi  \alpha)  f(\beta,\alpha) \,  \delta (x - \beta -\xi \alpha)  \, d\beta \, d\alpha
\nonumber \\
&=  \int_{\Omega}     F(\beta,\alpha) \,  \delta (x - \beta -\xi \alpha)  \, d\beta \, d\alpha  
\nonumber \\
&
+\xi   \int_{\Omega}    \frac{\alpha F(\beta,\alpha)}{\beta} 
\Big [  \delta (x - \beta -\xi \alpha) - \delta (x  -\xi \alpha) \Big ] 
 \, d\beta \, d\alpha  \nonumber \\
&
 +{\rm sgn}(\xi) D(x/\xi)
 \nonumber \\
&
\equiv F_{DD}(x,\xi) + \xi F^1_+ (x, \xi)  +{\rm sgn}(\xi) D(x/\xi) \  , 
\label{GPDf}
\end{align}
where
  \begin{align}
    F_{DD} (x,\xi) = \int_{\Omega}  F (\beta,\alpha)
    \,  \delta (x - \beta -\xi \alpha)    \, d\beta \, d\alpha
\label{FDDpi}
\end{align}
is GPD constructed from 
DD $ F (\beta,\alpha)$
by the same formula as in scalar case.
Another term 
\begin{align}
{F^1_+ (x,\xi)} \equiv   \int_{\Omega}  
                        \left ( \frac{\alpha}{\beta}\,  F (\beta,\alpha) \right )_+ \, 
                                     \delta (x - \beta -\xi \alpha) \, d\beta \, d\alpha
 \end{align}
is  a GPD built from the ``plus'' part of the DD $\alpha F(\beta,\alpha)/\beta$.
The latter, of course, may be written as
$G(\beta, \alpha)$, but in  the 
spirit of the one-DD formulation,
one may wish  to express the results in terms 
of just one function $F(\beta,\alpha)$.

\section{Factorized DD Ansatz}

In the forward limit  $\xi=0$, GPD  $H(x,\xi)$ converts into the usual 
parton distribution $f(x)$. In the 
one- DD formulation, we may write
\begin{align}
f(x) &=
 x     \int_{-1+|x|} ^{1  -|x|}  f(x,\alpha) \,   d\alpha   \   .
\label{GPDtofx}
\end{align}
Thus,  the forward  distributions $f(x)$  are  obtained by integrating 
over vertical lines $\beta =x$  in the $(\beta,\alpha)$ plane.
For nonzero $\xi$, GPDs  are obtained from  DDs  through  integrating them 
along   the lines $\beta=x-\xi \alpha$ having   $1/\xi$ slope. 
The reduction formula
  (\ref{GPDtofx}) 
suggests   the {\it factorized DD Ansatz} 
\begin{align}
f(\beta,\alpha) =  h(\beta,\alpha) \, {f(\beta)}/{\beta}  \   ,
\label{FDDA}
\end{align}
where $f(\beta)$   is the forward  distribution, while  $h(\beta,\alpha) $  determines DD profile in the 
$\alpha$  direction  and satisfies the normalization condition
\begin{align}
 \int_{-1+|\beta |} ^{1  -|\beta|}     h(\beta,\alpha) \,   d\alpha =1\   .
\label{hnorm}
\end{align}

 The profile  function should be symmetric with respect to
$\alpha \to -\alpha$ because 
DDs $ f(\beta,\alpha)$ 
are even in $\alpha$ \cite{Mankiewicz:1997uy,Radyushkin:1998bz}.
For a fixed $\beta$, the function  $  h(\beta,\alpha) $ 
describes how   the longitudinal momentum transfer $r^+$
is shared between the two partons.  Hence,  it is natural to expect that 
 the shape of $h(\beta,\alpha)$   
should  look like a symmetric meson 
distribution amplitude (DA) $\varphi (\alpha)$.  
Since  DDs have the support restricted by  
 $|\alpha| \leq 1- |\beta| $, to get a 
more complete  analogy
with DAs, 
it makes sense to rescale $\alpha$ as $\alpha = (1-|\beta|)  \gamma$
introducing  the  variable $\gamma$ with $\beta$-independent limits:
$-1 \leq \gamma \leq 1$. 
 The simplest model is to assume 
that the $\gamma$--profile  is  
 a  universal function  $g(\gamma)$ for all $\beta$. 
Possible simple choices for  $g(\gamma)$ may be  $\delta(\gamma)$
(no spread in $\gamma$-direction),  $\frac34(1-\gamma^2)$
(characteristic shape for asymptotic limit 
of nonsinglet quark distribution amplitudes), 
 $\frac{15}{16}(1-\gamma^2)^2$
(asymptotic shape of gluon distribution amplitudes), etc.
In the variables $\beta,\alpha $, these models can be treated as specific
 cases of the general profile function 
 \begin{equation}
 h^{(N)}(\beta,\alpha) = \frac{\Gamma (2N+2)}{2^{2N+1} \Gamma^2 (N+1)}
\frac{[(1- |\beta|)^2 - \alpha^2]^N}{(1-|\beta|)^{2N+1}} \,  , \label{modn} 
 \end{equation}
whose width is governed by the parameter $N$. 

\begin{figure}[ht]
   \begin{center} 
   \includegraphics[scale=0.18]{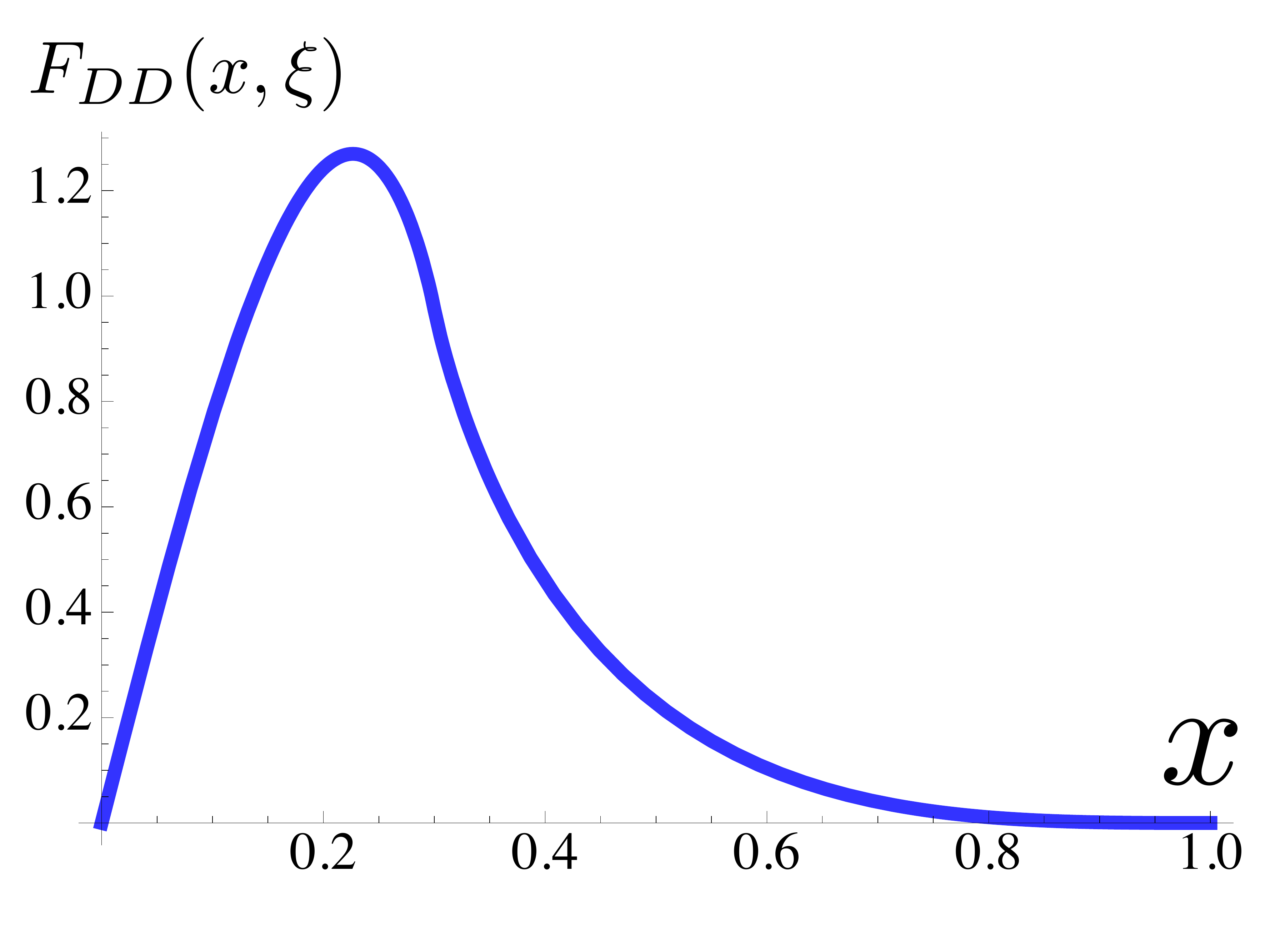}   
   \end{center}
   \caption{ GPD  $F_{DD} (x,\xi)$    for $\xi=0.3$.}
   \label{fig:FDD}
   \end{figure}

To give a graphical example, 
we show in Fig.\ref{fig:FDD} 
the simplest  GPD 
$F_{DD} (x,\xi)$ (\ref{FDDpi}) built from 
the model 
\begin{align}
F(\beta,\alpha) = {f(\beta)} 
 h^{(1)}(\beta,\alpha) 
\label{FDDmod}
\end{align}
with forward distribution
\begin{align}
f^{\rm mod} (x)=(1-x)^3/\sqrt{x}
\end{align}
and $N=1$ profile function
(analytic results for $f^{\rm mod} (x)=(1-x)^3 x^{-a}$ and $N=1$  profile
may be found in Refs. \cite{Musatov:1999xp,Radyushkin:2000uy,Belitsky:2005qn}).
   The model forward function was chosen
   in the form reproducing the 
   $x \to 1$ behavior of the nucleon 
   parton distributions and the 
   $\sim x^{-0.5}$  Regge behavior 
   of valence part of quark distributions 
   for small $x$, which was 
   taken for simplicity,  though the GPD shown corresponds 
   to the $C$-even component (the full function is antisymmetric in
   $x $, and only the $x\geq 0$ part is shown).

\section{Nucleon GPDs}

\subsection{Definitions of DDs and GPDs}

In the  nucleon case, for unpolarized target,   one can parametrize  
 \begin{align}
&  \langle p' |   \bar \psi(-z/2) \slashed z   \, \psi (z/2)|p \rangle  |_{\rm twist-2} 
\label{pmurmu2}
 \\ & = 
\int_{\Omega}   e^{-i \beta (Pz) -i\alpha (rz)/2} \,  \Big [ 
\bar u ( p')  \slashed z   \,  u ( p)\, a  (\beta, \alpha) \nonumber
 \\  &
+ \frac{\bar u ( p')     u ( p) }{2 M_N} \, \big [ 2\beta (Pz) +\alpha (rz) \big ]   b  (\beta, \alpha)   
\Big ]  \, d\beta \, d\alpha \ 
+{\cal O} (z^2)  
 \  .  \nonumber
\end{align}
Here, the functions $a,b$ are  DDs corresponding to the 
combinations $A=H+E$ and $B= -E$
of usual  GPDs $H$ and $E$ (see Ref. \cite{Belitsky:2005qn}).
These GPDs  may be expressed in terms of  relevant DDs as
\begin{align}
 {A(x,\xi) }  =
 \int_{\Omega}    a(\beta,\alpha) \,  \delta (x - \beta -\xi \alpha)  \, d\beta \, d\alpha 
\label{GPDA}
\end{align}
and
\begin{align}
 {B(x,\xi) }  ={x}
 \int_{\Omega}    b(\beta,\alpha) \,  \delta (x - \beta -\xi \alpha)  \, d\beta \, d\alpha  \  .
\label{GPDB}
\end{align}

Notice that we have two different types of relations between
GPDs and DDs: $A(x, \xi)$ is obtained from
its DD $a(\beta,\alpha)$ just like in the
simplest scalar case, while 
$B(x, \xi)$ is calculated from
 $b(\beta,\alpha)$  using the formula
 with the one-DD representation structure.
 The difference, of course, is due to
 the factor $[ 2\beta (Pz) +\alpha (rz)  ] $
 in the $b$-part.

In the forward limit, we have
\begin{align}
A(x,0) =H(x,0)+E(x,0) = f(x) +e(x)
\end{align}
and 
\begin{align}
B(x,0) =-E(x,0) = -e(x)  \ .
\end{align}
These reduction formulas suggest the model  representation
\begin{align}
a(\beta,\alpha) = 
f(\beta,\alpha) + e (\beta,\alpha) 
\end{align}
and 
\begin{align}
b(\beta,\alpha) = 
- \frac{e(\beta,\alpha)}{\beta} \  .
\end{align}
Because of possible singularity of  $e(\beta,\alpha)/\beta$  at $\beta=0$,
we write it in the ``DD$_++ D$''  representation:
\begin{align}
b(\beta,\alpha) = - \left ( \frac{e(\beta,\alpha)}{\beta}  \right )_+  
+
\delta (\beta) \frac{D(\alpha) }{\alpha}  \  ,
\end{align}
where $D(\alpha)$ is  the $D$-term.

\subsection{General results for GPDs}
As a result, we have
\begin{align}
& {H(x,\xi) }  =  {A(x,\xi) }  + {B(x,\xi) }  \\ &= 
 \int_{\Omega}   [ f(\beta,\alpha) + e(\beta,\alpha) ]
 \,  \delta (x - \beta -\xi \alpha)  \, d\beta \, d\alpha \nonumber \\ & 
 -  
  x \int_{\Omega}   \left [ \left ( \frac{e(\beta,\alpha)}{\beta}  \right )_+ -
  \delta (\beta) \frac{D(\alpha) }{\alpha} \right ]
  \,  \delta (x - \beta -\xi \alpha)  \, d\beta \, d\alpha \nonumber \\ &
  = F_{DD} (x,\xi) +  E_{DD}  (x,\xi)  -  E_+(x,\xi)  +{\rm sgn} (\xi) \,    D(x/\xi)  \ , \nonumber 
  \end{align}
  where 
  \begin{align}
    F_{DD} (x,\xi) = \int_{\Omega}  f (\beta,\alpha)
    \,  \delta (x - \beta -\xi \alpha)    \, d\beta \, d\alpha
\label{GPDHf}
\end{align}
and 
 \begin{align}
    E_{DD}  (x,\xi) = \int_{\Omega}  e (\beta,\alpha)
    \,  \delta (x - \beta -\xi \alpha)    \, d\beta \, d\alpha
\label{GPDE}
\end{align}
are build from DDs by   simplest  formulas  not involving a  division by $\beta$ factors, while 
  \begin{align}
  \frac{E_+ (x,\xi)}{  x } =&  \int_{\Omega}  \frac{e(\beta,\alpha)}{\beta}     \Big [ 
    \,  \delta (x - \beta -\xi \alpha)    - \delta (x -\xi \alpha)      \Big ]  \, d\beta \, d\alpha  
    \nonumber \\ &= 
      \int_{\Omega}  \left( \frac{e(\beta,\alpha)}{\beta}  \right )_+    
          \delta (x - \beta -\xi \alpha)   \, d\beta \, d\alpha  
\label{GPDE+}
\end{align}
has the  structure  of  a one-DD representation.  Since 
$E_+ (x,\xi)/  x $ is built from the 
``plus'' part of a DD it should satisfy 
\begin{align}
\int_{-1}^1 \, E_+(x,\xi) \, dx = \int_{\Omega}  
                           \left [  \frac{e (\beta,\alpha)}{\beta}  \right ]_+ \, 
                                    \, d\beta \, d\alpha = 0 \ 
 . 
\label{H+zero}
\end{align}

 \begin{figure}[t]
 \begin{center} 
 \includegraphics[scale=0.2]{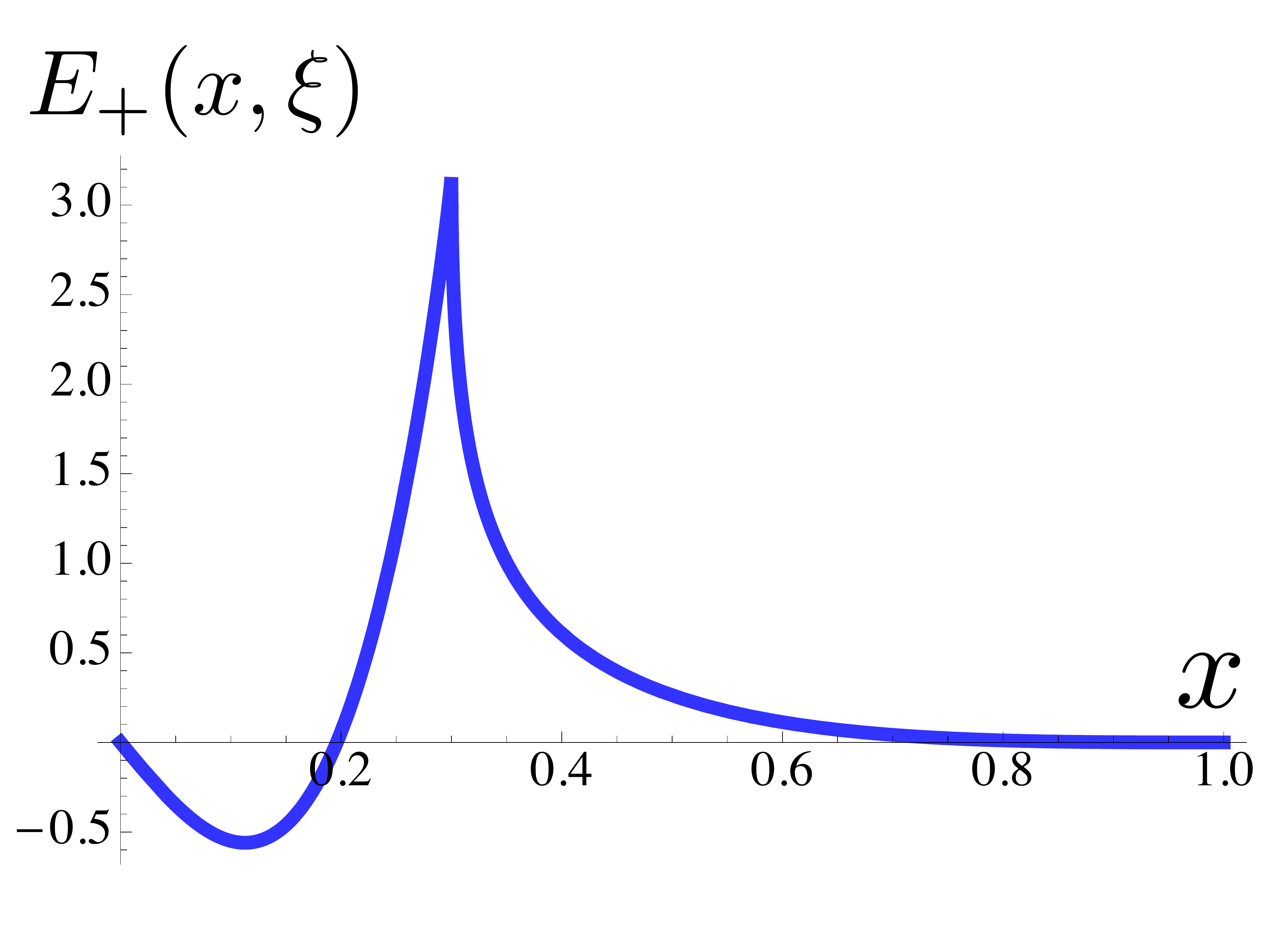}   
 \end{center}
 \caption{ GPD  $E_+ (x,\xi)$    for $\xi=0.3$.}
 \label{fig:Eplus}
 \end{figure}

 Being (for $C$-even combination) an even function of $x$,  the function $E_+ (x,\xi)/x$ obeys 
\begin{align}
\int_0^1   \frac{E_+ (x,\xi)}{  x } \, dx =0 \ .
\end{align}

\subsection{Modeling GPDs}

\begin{figure}[b]
 \begin{center} 
 \includegraphics[scale=0.25]{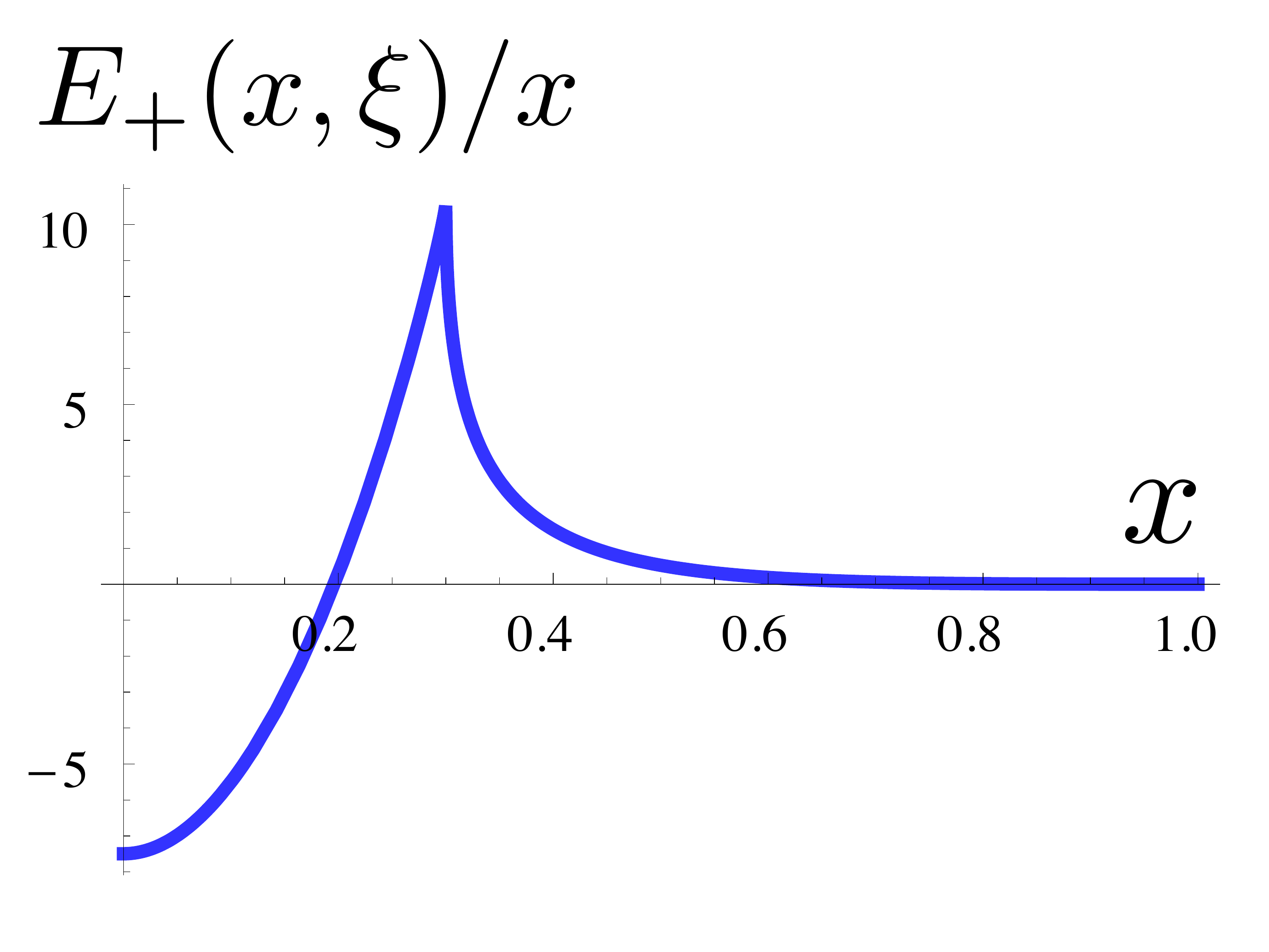}   
 \end{center}
 \caption{ GPD  $E_+ (x,\xi)/x$    for $\xi=0.3$.}
 \label{fig:Eplus_overx}
 \end{figure}

To illustrate the structure of  $E_+ (x,\xi)$, 
we show it in Fig.\ref{fig:Eplus} 
using the model based on  
\begin{align}
e(\beta,\alpha) = {e(\beta)} 
 h^{(1)}(\beta,\alpha) 
\end{align}
with $N=1$ profile function  and the same forward distribution
$e(x)=(1-x)^3/\sqrt{x}$ that was used 
to model $F_{DD}$ above.
Again, we have in mind the $C$-even,
quark+antiquark part of the distribution,
and valence-type functional form
is used to simplify the illustration.
One can see that $E_+ (x,\xi)$
is a regular  function, and vanishing of 
$E_+ (x,\xi)/x$ integral is due to compensation
over positive and negative parts
(see Fig.\ref{fig:Eplus_overx}) 
rather than because of subtraction
of a $\delta (x)$  term.

In a more realistic modeling,
one should adjust normalization of 
$e(x)$ to reflect its relation
to the anomalous  magnetic moment.
Also, the fits of the nucleon elastic
form factors \cite{Guidal:2004nd} 
suggest  for $e(x)$ a higher power 
of $(1-x)$.
However, our  aim while  showing the curves in the present paper
is just to illustrate  the qualitative features 
of various GPD models,
so we will stick to the same generic
forward function  both for $f(x)$ and $e(x)$.
 
The function $E_+ (x,\xi)$ may be displayed as 
 \begin{align}
&  E_+ (x,\xi) = x \int_{\Omega}  
   \frac{ e (\beta,\alpha)}{\beta} \left [ 
    \,  \delta (x - \beta -\xi \alpha)  -  \delta (x  -\xi \alpha) 
     \right ] \, d\beta \, d\alpha
     \nonumber \\ &
       =\int_{\Omega}  e (\beta,\alpha)
           \,  \delta (x - \beta -\xi \alpha)    \, d\beta \, d\alpha
       \nonumber \\ &    
       +\xi  \int_{\Omega}  
           \frac{\alpha}{\beta}\,  e (\beta,\alpha) \, 
           \left [  
             \delta (x - \beta -\xi \alpha)  -   \delta (x  -\xi \alpha) 
            \right ] \, d\beta \, d\alpha
              \nonumber \\ &
                   =E_{DD} (x,\xi) +\xi  \int_{\Omega}  
                       \left ( \frac{\alpha}{\beta}\,  e (\beta,\alpha) \right )_+ \, 
                                                \delta (x - \beta -\xi \alpha) \, d\beta \, d\alpha     
                                                \nonumber \\ &                                          
                                                \equiv E_{DD} (x,\xi)  +{\xi } E^1_+ (x,\xi)
 \  ,
\label{GPDE1}
\end{align}
where
\begin{align}
{E^1_+ (x,\xi)} \equiv   \int_{\Omega}  
                        \left ( \frac{\alpha}{\beta}\,  e (\beta,\alpha) \right )_+ \, 
                                     \delta (x - \beta -\xi \alpha) \, d\beta \, d\alpha \ .
 \end{align}
Since $E^1_+(x,\xi) $ is built from
the ``plus'' part of a DD, its $x$-integral
from $-1$ to 1 is equal to zero, but 
in fact it vanishes also for a simpler reason
that $E^1_+(x,\xi) $  is an odd function of $x$.
So, in this case, we cannot make any conclusions about the  magnitude of the 
$x$-integral of $E^1_+(x,\xi) $ 
from  0 to 1. 

\begin{figure}[ht]
\begin{center} 
\includegraphics[scale=0.2]{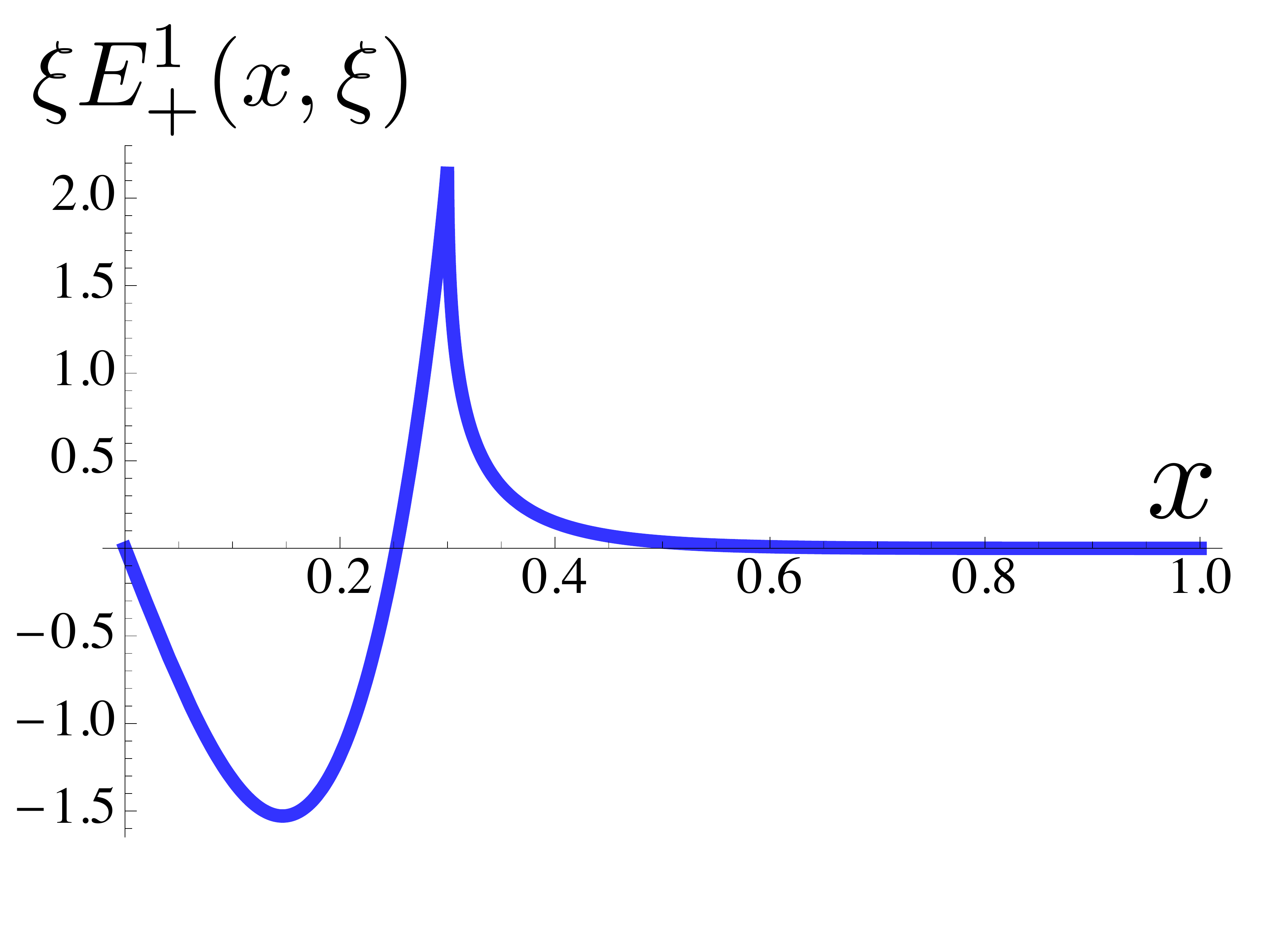}   
\end{center}
\caption{Function  $\xi E^1_+ (x,\xi) $  for $\xi=0.3$.}
\end{figure}

Summarizing,  GPD $E_+$ is  obtained
from the naive $E_{DD}$ function by
{\it adding} to it the $\xi E^1_+(x,\xi) $
term, which results in a rather 
nontrivial non-monotonic behavior of the
$E_+$ 
function. To get the full  GPD $E$,
one should subtract also the $D$-term
contribution:
 \begin{align}
  {E(x,\xi) } =& E_+ (x,\xi) -  {\rm sgn} (\xi )    D(x/\xi)
 \\ & =  E_{DD}  (x,\xi) + \xi   E^1_+(x,\xi)   - {\rm sgn} (\xi )    D(x/\xi)  \ .
   \nonumber
 \end{align}

For GPD   $H$,  we then have
\begin{align}
 {H(x,\xi) } =  F_{DD}  (x,\xi) - \xi   E^1_+(x,\xi)   + {\rm sgn} (\xi )    D(x/\xi)  \ .
\end{align}
Now one should {\it subtract}
$\xi  E^1_+(x,\xi) $  from the naive
$F_{DD}$ function and then add the 
$D$-term  contribution.

  \begin{figure}[htb]
  \begin{center} 
  \includegraphics[scale=0.2]{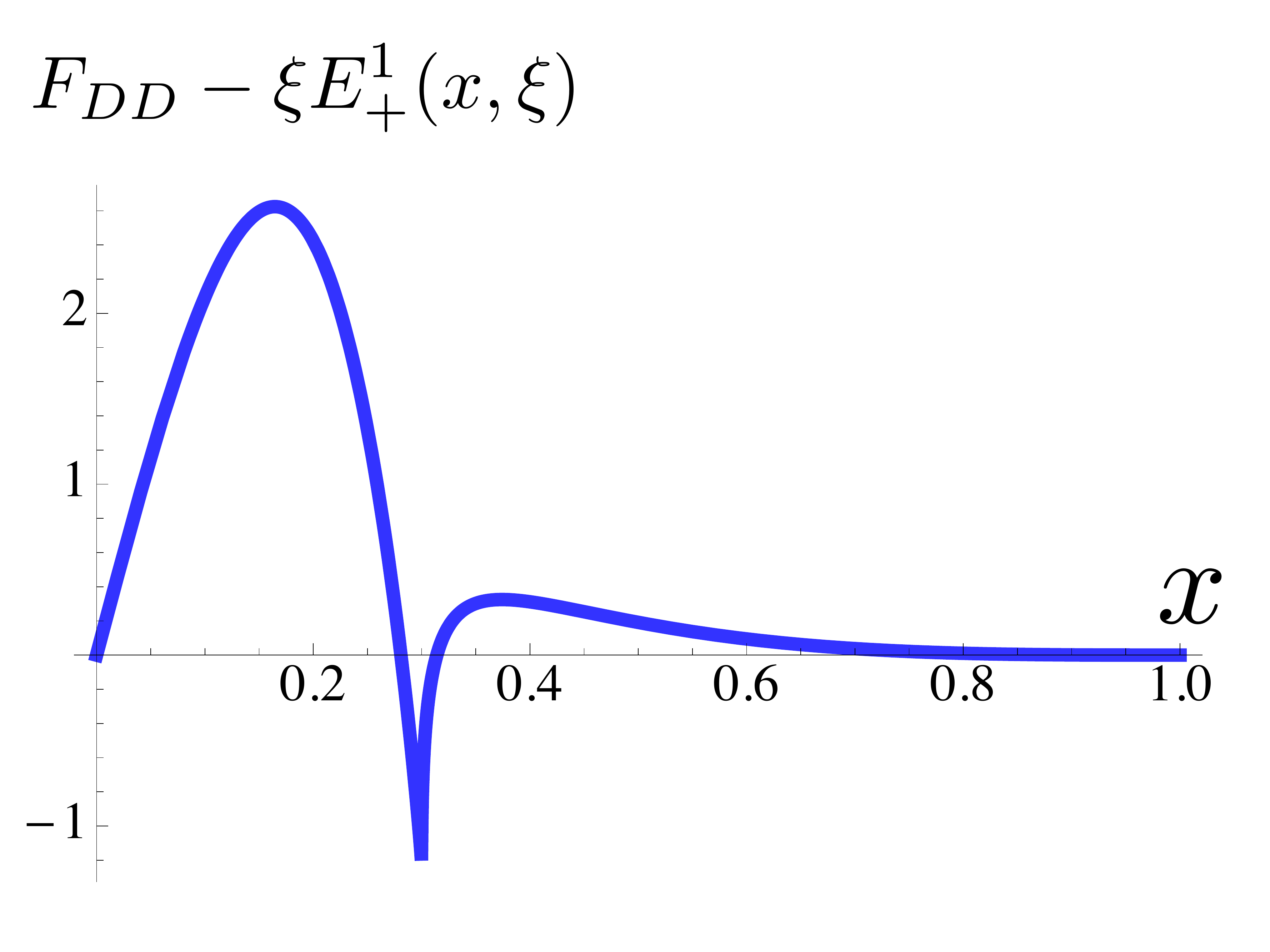}   
  \end{center}
  \caption{ Model nucleon GPD $H(x,\xi)$ (without $D$-term)
      for $\xi=0.3$.}
  \end{figure}

 Comparing this result with the pion case  for which 
 \begin{align}
& H(x,\xi) =   F_{DD}(x,\xi) + \xi F^1_+ (x, \xi)  +{\rm sgn}(\xi) D(x/\xi) \  , 
\label{GPDf}
\end{align}
 we see that the structure of the pion GPD
 $H_+$ is similar to that of the
 nucleon GPD $E_+$: the term 
 $\xi F^1_+ (x, \xi) $ is added to
 $F_{DD}(x,\xi)$  rather than subtracted.
 However, in case of the   nucleon 
 GPD $H$, 
 the extra term
 is built from the second nucleon DD $e(\beta,\alpha)$ rather than from   $f(\beta,\alpha)$,  and it is subtracted from 
 $F_{DD}(x,\xi)$ 
 rather than added to it.

 \subsection{Polynomiality} 
 
 Taking the  $x^n$ moment of $H(x,\xi)$ in this construction, we note 
 that the $F_{DD} (x,\xi)$ term produces only the powers of 
 $\xi$ up to $\xi^n$.  Next observation is   that   the highest, namely $n^{\rm th}$ power of $\xi$ in the 
 $x^n$ moment of $E^1_+ (x,\xi)$ involves  the integral
\begin{align}
  \int_{\Omega}  
                          \alpha^n \,  \left ( \frac{\alpha}{\beta}\,  e (\beta,\alpha) \right )_+ \, 
                                    \, d\beta \, d\alpha
 \end{align}
 that vanishes because the integrand is a ``plus'' distribution 
 with respect to $\beta$.  Hence, $\xi E^1_+ (x,\xi)$
 term  also cannot produce the $\xi^{n+1}$ contribution for the 
 $x^n$ moment of $H(x,\xi)$. Such a term is produced by the 
 $D$-term only.
 
\subsection{Comparison with ``DD plus D-term''  model}

The usual ``DD plus D-term'' model in the context of the 
present paper
corresponds to ``$F_{DD}$ plus D-term''  combination,
i.e. modeling nucleon GPDs  without subtracting the 
 $\xi   E^1_+(x,\xi)$ term when modeling $H(x,\xi)$,  
 (or  adding it when modeling $E(x,\xi)$).
 
 In a sense, our new model results from the 
 old ``DD plus D'' model by substituting 
 $ {\rm sgn}(\xi) D(x/\xi)$ with  $-\xi   E^1_+(x,\xi)+ {\rm sgn}(\xi) D(x/\xi)$.   
 
 Since the $D$-term is fitted to data,
 one may wonder if  adding  $\xi   E^1_+(x,\xi)$ 
 may be absorbed by redefinition of the $D$-term.
 However, there are  important 
 qualitative differences  between $   E^1_+(x,\xi)$
  and $ D(x/\xi)$.
 First, 
 the support region of 
 $   E^1_+(x,\xi)$ is not restricted to the segment 
 $|x| \leq \xi$. Furthermore, existing  
 models of $D(x/\xi)$ assume that it is a continuous 
 function that vanishes 
 not only outside the central $|x| \leq \xi$ region, but also 
 at the border points $|x|=\xi$ 
 (otherwise, GPDs $H$ and $E$ would be discontinuous 
 at the border points, and pQCD  factorization formula 
 for DVCS would make no sense).
 As we have seen, $   E^1_+(x,\xi)$ is a continuous function
 of $x$ in the whole $|x| \leq 1$ region, and it is not
 vanishing at the border points $|x |=\xi$.
 
 Thus, the most apparent  difference between  the two models
 is that the value of $H(\xi,\xi)$, the GPD at the 
 border point, in the new model  is different from that given 
 by GPD $F_{DD} (\xi, \xi)$ built solely from DD $f(\beta,\alpha)$
 related to the usual forward parton density $f(\beta)$.
 Furthermore, this difference is determined 
 by DD $e(\beta, \alpha)$  that is  related to GPD $E(x,\xi)$ 
 invisible in the forward limit.

 \section{Summary} 

Summarizing, the model for GPD $H$ proposed in this paper 
differs from the ``old-fashioned'' DD+D model by
an extra $-\xi E^1_+ (x,\xi)$ term  constructed 
from the DD $e(\beta,\alpha)$ corresponding to the GPD 
$E (x,\xi)$.
The inclusion of such a  term  modifies the original DD-based term
$F_{DD} (x,\xi)$ at the border points
$|x|=\xi$ and outside  the central $|x/\xi| \leq 1$ region, which 
may have strong phenomenological consequences. 

\section*{Acknowledgements}

I  thank 
    H. Moutarde and A. Tandogan 
for  discussions,  and C. Mezrag for correspondence.

This work is supported by Jefferson Science Associates,
 LLC under  U.S. DOE Contract No. DE-AC05-06OR23177.


 \bibliographystyle{apsrev4-1.bst}
\bibliography{nucleon0514.bib}

 \end{document}